\documentclass{article}

\usepackage[utf8]{inputenc}
\usepackage[a4paper]{geometry}
\usepackage{amsmath}
\usepackage{amssymb}
\usepackage{multirow}
\usepackage{authblk} 
\usepackage{pgfplots}
\pgfplotsset{compat=1.14}
\usetikzlibrary{shapes.symbols,graphs}

\usepackage[sorting=none,style=numeric-comp]{biblatex}
\usepackage[pdfauthor={Filipe F. C. Silva, Pedro M. S. Carvalho, Luis A. F. M. Ferreira and Yasser Omar},
            pdftitle={A QUBO Formulation for Minimum Loss Spanning Tree Reconfiguration Problems in Electric Power Networks},
            hidelinks,pdftex]{hyperref}
            
\usepackage{caption}

\title{A QUBO Formulation for Minimum Loss Spanning Tree Reconfiguration Problems in Electric Power Networks}

\author[1,2,3]{Filipe F. C. Silva}
\author[1,3]{Pedro M. S. Carvalho}
\author[1,3]{\\Luís A. F. M. Ferreira}
\author[2,3,4]{Yasser Omar}
\affil[1]{{\small INESC-ID, Sustainable Power Systems Group, Portugal}}
\affil[2]{{\small Instituto de Telecomunicações, Physics of Information and Quantum Technologies Group, Portugal}}
\affil[3]{{\small Instituto Superior Técnico, University of Lisbon, Portugal}}
\affil[4]{{\small Portuguese Quantum Institute, Portugal}}

\date{March 2022}

\begin{document}

\maketitle

\begin{abstract}
We introduce a novel quadratic unconstrained binary optimization (QUBO) formulation for a classical problem in electrical engineering -- the optimal reconfiguration of distribution grids.
For a given graph representing the grid infrastructure and known nodal loads, the problem consists in finding the spanning tree that minimizes the total link ohmic losses.
A set of constraints is initially defined to impose topologically valid solutions.
These constraints are then converted to a QUBO model as penalty terms.
The electrical losses terms are finally added to the model as the objective function to minimize.
In order to maximize the performance of solution searching with classical solvers, with hybrid quantum-classical solvers and with quantum annealers, our QUBO formulation has the goal of being very efficient in terms of variables usage.
A standard 33-node test network is used as an illustrative example of our general formulation.
Model metrics for this example are presented and discussed.
Finally, the optimal solution for this example was obtained and validated through comparison with the optimal solution from an independent method.
\end{abstract}

\section{Introduction}
Electrical distribution grids are facing the fundamental paradigm shift of becoming low-carbon smart grids.
Under this scenario, the grids must be able to accept an increasing amount of distributed renewable energy generation (e.g., from urban and residential photovoltaic panels and small wind turbines) and of electric vehicles load.
Handling the volatile nature of new load and generation poses new challenges to the grid operation, since the network must be regularly reconfigured in order to operate with minimum energy losses \cite{Silva2021}.
Those reconfigurations consist of switching on and off several network links.
As the network reacts to the constant changes on load and generation profiles, the reconfiguration decisions need to be fast and optimal in order to be effective.

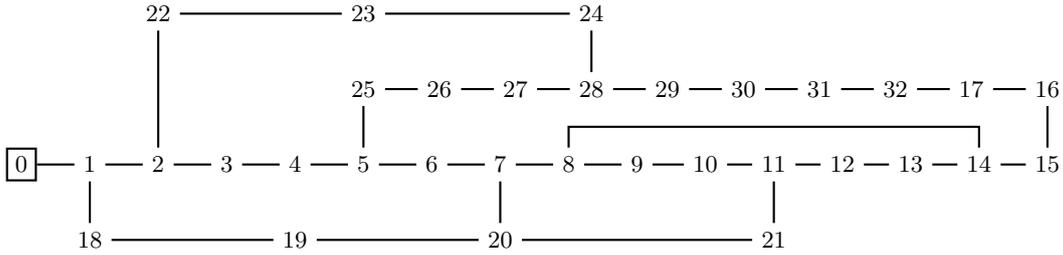
\begin{figure}
	\centering
	\small
	\begin{tikzpicture} [thick,xscale=.9]
		\graph [no placement] {
			0[draw] -- 1[x=1] -- 2[x=2] -- 3[x=3] -- 4[x=4] -- 5[x=5] -- 6[x=6] -- 7[x=7] -- 8[x=8] -- 9[x=9] -- 10[x=10] -- 11[x=11] -- 
				12[x=12] -- 13[x=13] -- 14[x=14] -- 15[x=15];
			1 -- {[y=-1] 18[x=1] -- 19[x=4] -- 20[x=7] -- 21[x=11]} -- 11;
			5 -- {[y=1] 25[x=5] -- 26[x=5+10/9] -- 27[x=5+20/9] -- 28[x=5+30/9] -- 29[x=5+40/9] -- 30[x=5+50/9] -- 31[x=5+60/9] --
				32[x=5+70/9] -- 17[x=5+80/9] -- 16[x=15]} -- 15;
			2 -- {[y=2] 22[x=2] -- 23[x=5] -- 24[x=5+30/9]} -- 28;
			7 -- 20;
		};
		\draw (8) |- (10,.5) -| (14);
	\end{tikzpicture}
	\caption{The standard 33-node test network. The substation is node 0.}
	\label{fig:net33}
\end{figure}

In this paper, the distribution grid is represented through a network graph where one of the nodes is the substation and the remaining nodes are the network loads.
The network links are the electrical lines connecting the nodes.
Figure \ref{fig:net33} represents the 33-node test network used in this paper to illustrate our formulation.
This network was defined in \cite{Wu1989} and it is widely used in power systems literature to benchmark network optimization algorithms \cite{Mishra2017}.
Besides the topology, this network model includes the electrical characteristics of the links (longitudinal impedance) and the electrical power taken on each load.

Although this network contains several cycles, a distribution grid is always operated radially (i.e., with no cycles allowed) \cite{Carvalho2018}.
Thus, several links must be switched off (or opened) in order to obtain a spanning tree connecting all network nodes.
Given that all network links are susceptible to switching, there are 50751 distinct spanning trees (i.e., valid operating configurations) on this network.
Since this is a small search space, the optimal configuration can be easily found by exhaustive search with an efficient algorithm to generate all spanning trees.
Literature already reports several methods for the minimum loss reconfiguration problem \cite{Mishra2017}.
The first of these reports dates back to 1975 with a heuristic branch-and-bound sequential switch opening method \cite{merlin1975}.

The contribution of this paper is the development of a novel Quadratic Unconstrained Binary Optimization (QUBO) model \cite{Glover2019,Glover2017} for the minimum loss reconfiguration problem.
Formulating this problem as QUBO enables the use of quantum annealers \cite{Hauke_2020,nasa2017,Albash_2018} such as the 5000-qubit D-Wave Advantage system \cite{dwave_advantage} and classical-quantum hybrid solvers such as the D-Wave Hybrid Solver Service \cite{dwave_hybrid} to find faster and better decisions for network reconfiguration.
An important feature of our QUBO formulation is that of being very efficient in terms of QUBO variables count and product terms count such that network models of interesting size can be fit into size-limited quantum annealers.

Some QUBO formulations of tree problems can be found in the literature \cite{Lucas_2014,fowler}, although those general formulations are not as efficient as the one presented in this paper because they do not take into account the specific nature of the electrical network topology, namely in terms of sparsity.
Additionally, we are not aware of any other QUBO formulation of spanning tree problems for minimum quadratic cost over the network flows, as is the case of minimum electrical loss problems.

Several steps are needed in order to build a complete QUBO model for solving our optimization problem:
\begin{enumerate}
	\item Express topological constraints as a Constraint Satisfaction Problem (CSP).
	These constrains ensure that only valid spanning trees are returned.
	\item Define auxiliary variables as additional CSP constraints.
	These variables are later needed for the electrical losses model.
	\item Convert CSP to a QUBO model.
	Each CSP constraint is converted to a QUBO expression which penalizes (i.e., assigns a higher cost to) any solution violating the constraint.
	The QUBO model is the sum of all constraint expressions.
	An optimal solution for this model complies with all CSP constraints, thus being a valid solution.
	\item Add electrical losses terms to the QUBO model.
	The optimal solution for the complete QUBO model is the one minimizing electrical losses while still being topologically valid.
\end{enumerate}

This paper is organized as follows.
After this Introduction, the optimization problem is described in Section \ref{section:description}.
Afterwards, in Section \ref{sec:constraints}, the problem constraints are formulated and, in Section \ref{section:formulation}, the method of obtaining the QUBO model is described.
Then, in Section \ref{section:results}, metrics and validation of the QUBO model are presented and discussed.
Finally, Section \ref{section:conclusions} concludes the paper.

Throughout the paper, the terms \emph{network}, \emph{node} and \emph{link} are used when node and link intrinsic attributes are important in the context (such as node load and link resistance in an electrical network), and the terms \emph{graph}, \emph{vertex}, \emph{edge} and \emph{arc} are used when the focus is on graph theory or topology.

\section{Problem Description}
\label{section:description}
The model of an electrical network includes the connectivity graph $G=(V,E)$, the complex phasor of nodal electrical load current\footnote{Electrical loads are typically modeled as constant power (PQ loads), but we are modeling them as constant current loads since this allows a significant simplification to our losses model without changing the optimal configuration for the 33-node network. We are also assuming that voltage angle is zero across the network.}
$I^L_n$ and link electrical resistance\footnote{Since load current does not depend on load voltage and we are only concerned with active power losses, longitudinal link reactance can be ignored. Also, the 33-node network model does not specify transversal link susceptance.}
$R_{uv}$, with $n \in V$ and $(u,v) \in E$.
Given this model as the input, the problem goal is to find the spanning tree $T^*$ of $G$ which minimizes the sum of active power losses on each link of the tree, as defined by
\begin{equation}\label{eq:optim}
	T^* = \operatorname*{arg\,min}_{T \in \mathrm{ST}(G)} \sum_{(u,v) \in E(T)} L_{uv}(T)
\end{equation}
where $\mathrm{ST}(G)$ is the set of all spanning trees of $G$, $E(T)$ is the set of all links on $T$ and $L_{uv}(T)$ is the power losses function on link $(u,v)$ with tree $T$.
This function is given by
\begin{equation}\label{eq:losses}
	L_{uv}(T) = R_{uv}\left|I_{uv}(T)\right|^2
\end{equation}
where $I_{uv}(T)$ is the complex phasor of the electrical current flowing on link $(u,v)$ with tree $T$.
This current is given by
\begin{equation}\label{eq:loads}
	I_{uv}(T) = \sum_{n \in D_{uv}(T,v_0)} I^L_n
\end{equation}
where $D_{uv}(T,v_0)$ is the set of all nodes downward of link $(u,v)$ across $T$, i.e., the nodes with a path to the root node $v_0 \in V$ (the substation) along $T$ which includes the link $(u,v)$.
This set includes either $u$ or $v$ depending on $T$.
$D_{uv}(T,v_0)$ can be extended for the case $(u,v) \notin E(T)$ yielding the empty set.

\subsection{Problem partitioning into biconnected components}\label{sec:partitioning}
Let $D_m'(T,v_0)$ be the set of all nodes downward of node $m$ across $T$.
If, for a given tree $T$, $m$ is itself downward of a link $(u,v)$, then a node $n$ downward of $m$ is also downward of $(u,v)$, i.e., $n \in D_m'(T,v_0) \wedge m \in D_{uv}(T,v_0) \implies n \in D_{uv}(T,v_0)$.	
If $n$ is downward of $m$ in all possible spanning trees, i.e., if $n \in D_m'(T,v_0) \quad \forall T \in \mathrm{ST}(G)$, then $n$ is downward of $(u,v)$ iff $m$ is also downward of $(u,v)$ regardless of the spanning tree considered.
Under this assumption, if $n$ is removed from $G$ while its load current $I^L_n$ is added to $I^L_m$ then there is no change in the current $I_{uv}$ of any link $(u,v)$ not placed in the path between $m$ and $n$, since
\begin{equation}
	I_{uv} = \sum_{k \in D_{uv}(T,v_0)} I^L_k = \sum_{k \in D_{uv}(T,v_0)\setminus\{n\}} \begin{cases}
		I^L_m + I^L_n & \text{if} \ k = m\\
		I^L_k & \text{otherwise}
	\end{cases}\quad \forall T \in\mathrm{ST}(G)\,.
\end{equation}

\begin{figure}
	\centering
	\begin{tikzpicture}[thick]
		\node[draw] (v0) at (0,0) {$v_0$};
		\node (k) at (1,1) {$k$};
		\node (l) at (1,-1) {$l$};
		\node[draw,circle] (m) at (2,0) {$m$};
		\node (n1) at (3,1) {$n_1$};
		\node (n2) at (3,-1) {$n_2$};
		\node (n3) at (4,0) {$n_3$};
		
		\draw (v0) -- (k);
		\draw (v0) -- (l);
		\draw (k) -- (m);
		\draw (l) -- (m);
		\draw (m) -- (n1);
		\draw (m) -- (n2);
		\draw (n1) -- (n3);
		\draw (n2) -- (n3);
	\end{tikzpicture}
	\caption{With $v_0$ as the root, nodes $n_1$, $n_2$ and $n_3$ are downward of $m$ in all possible spanning trees of this network. Thus, the two components separated by $m$ (the biconnected components) can be optimized as independent problems.}
	\label{fig:biconn_comp}
\end{figure}
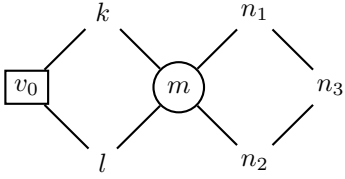

This procedure can be repeated with $m$ for all nodes in the same condition as $n$.
Also, from (\ref{eq:loads}) one can see that $I_{uv}$ is independent of the tree path from the root node to link $(u,v)$. 
Thus, more generally, if a node $m$ is a cutvertex separating $G$ into two of more connected components, each component can be optimized as a separate problem.
For the component including the root node, the total load of the other components is added to the load of $m$.
For the other components, $m$ is considered their root node.
If some of these components are trees, then there is no optimization to be done on those components.
Figure \ref{fig:biconn_comp} represents a network which can be partitioned into two components.

An equivalent statement can be made in terms of the biconnected components of $G$ -- each of these components can be treated as a separate optimization problem, excluding the trivial components consisting of a dyad (i.e., a pair of nodes with a link between them).
The graph of the 33-node test network (Figure \ref{fig:net33}) has two biconnected components: one is the dyad containing the link (0,1), and the other one contains the remaining links with node 0 removed.
The network to be optimized reduces to this second component, thus node 0 and link (0,1) can be removed and node 1 becomes the new root node.

\subsection{Spanning tree reduction through edge lifting}\label{sec:edge_lifting}
Let $G_C=(V_C,E_C)$ be a non-trivial biconnected component of $G$.
$G_C$ can be reduced to a topological minor $G_0=(V_0,E_0)$ where $V_0 \subset V_C$ through successive application of edge liftings -- the two edges incident to a vertex with degree 2 are replaced by a single edge incident to the two neighbors of that vertex while the vertex is removed \cite{diestel2000graph}.
This step is repeated for all vertices of $G_C$ with degree 2 with the exception of the root vertex $v_0 \in V_C$ (if it also has degree 2) since $v_0$ must also be in $V_0$.
Given that $G_C$ is a non-trivial biconnected component, its vertices have a degree of at least 2.
With the possible exception of the root vertex, the vertices of $G_0$ have a degree of at least 3 -- keeping the same degree as they have in $G_C$ -- since $G_0$ contains no vertices with degree 2.

As a result of the edge lifting transformation, each edge $(u,v) \in E_0$ represents a path of $G_C$ with one or more edges.
Let $P_{uv} \subset G_C$, with $V(P_{uv}) \cap V_0 = \{u,v\}$, be such a path and let $T_C$ be a spanning tree of $G_C$.
If all edges of $P_{uv}$ are in $E(T_C)$ (i.e., if $P_{uv} \subset T_C$), then the path is said to be closed in $T_C$; otherwise the path is open in $T_C$.
If a path is open, then exactly one of its edges is not in $E(T_C)$ and that edge is said to be the open edge of the path in $T_C$ -- if two or more edges in the path would not be in $E(T_C)$, then $T_C$ would be either a disconnected graph or not spanning to all vertices of $G_C$, and thus $T_C$ could not be a spanning tree of $G_C$.

Let $T_0$ be a spanning tree of $G_0$.
This tree corresponds to a $T_C$ such that any edge in $E(T_0)$ represents a closed path in $T_C$ and any edge in $E_0 \setminus E(T_0)$ represents an open path in $T_C$, i.e., $(u,v) \in E(T_0) \Leftrightarrow P_{uv}(G_C) \subset T_C$, $\forall (u,v) \in E_0$.
Thus, a given $T_0$ corresponds to all spanning trees of $G_C$ with a given set of open (or closed) paths -- the only difference between those trees is the set of open edges $Q = E_C \setminus E(T_C)$ belonging to those open paths.

With this method, the problem of finding the optimal spanning tree $T_C^*$ is equivalent to finding the optimal pair composed of the smaller spanning tree $T_0$ and the set $Q$ containing an open edge $q_{ab} \in E(P_{ab})$ for each edge $(a,b) \in E_0 \setminus E(T_0)$.
The new problem is then given by
\begin{equation}
	(T_0,Q)^* = \operatorname*{arg\,min}_{\substack{T_0 \in \mathrm{ST}(G_0)\\
													q_{ab} \in E(P_{ab})\ \forall (a,b) \in E_0 \setminus E(T_0)}}
												\sum_{(u,v) \in E(T_C(T_0,Q))} R_{uv} |I_{uv}(T_C(T_0,Q))|^2
\end{equation}
where $T_C$ is computed from the complete configuration $(T_0,Q)$ and $I_{uv}(T_C)$ is defined as in (\ref{eq:loads}).

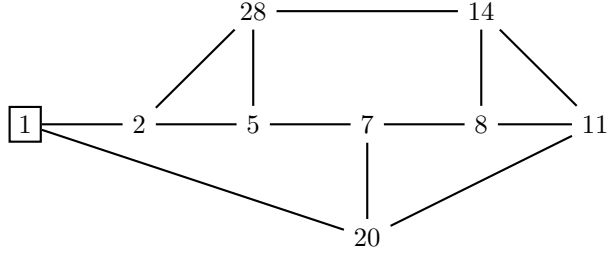
\begin{figure}
	\centering
	\begin{tikzpicture} [thick,scale=1.5]
		\graph [no placement] {
			1[draw] -- 2[x=1] -- 5[x=2] -- 7[x=3] -- 8[x=4] -- 11[x=5];
			1 -- 20[x=3,y=-1] -- 11;
			2 -- {[y=1] 28[x=2] -- 14[x=4]} -- 11;
			5 -- 28;
			7 -- 20;
			8 -- 14;
		};
	\end{tikzpicture}
	\caption{The reduced main biconnected component of the 33-node test network. The component root is vertex 1.}
	\label{fig:comp_reduced}
\end{figure}

As most of the nodes of the main biconnected component of the 33-node test network (figure \ref{fig:net33}) have degree 2, this reduction yields a much smaller topological minor, as shown in figure \ref{fig:comp_reduced}.
This reduced component has 463 distinct spanning trees, which compares with the 50571 spanning trees of the original network.

\section{Problem Constraints}\label{sec:constraints}
This section expresses the problem constraints in terms of binary variables (assuming the values 0 or 1) which will be part of the final QUBO model.
These constraints fall into two main categories: topology constraints, which impose that the problem solution is a spanning tree, and auxiliary variables constraints, which assign values to auxiliary variables needed for the QUBO cost function (in this case, to express the network power losses).
The first category encompasses vertex, edge, cycle and path constraints.

For the formulation of the problem, it is necessary to assign a direction to the edges of the spanning tree.
This direction is defined as pointing downwards, i.e., away from the root vertex of the tree.
The resulting directed spanning tree is thus a spanning arborescence composed of arcs (i.e., directed edges).

Let $G_0=(V_0,E_0)$ be the undirected graph of a reduced non-trivial biconnected component of the original network graph (as described in \ref{sec:edge_lifting}) and let $v_0 \in V_0$ be the root of $G_0$.
Let $G_D=(V_0,A)$ be the directed graph containing an arc for each valid direction of each edge of $G_0$.
Given our definition for edges direction, the edges incident to the root have only one valid direction -- away from the root.
Thus, each of these edges corresponds to a single arc on $A$, and since the root has no incoming arc, it is also the source of $G_D$.
All other edges admit the two directions depending on the spanning tree, thus assigning two arcs on $A$ with opposite directions for each edge.
The set of valid arcs for the undirected edge $(u,v) \in E_0$ is then given by
\begin{equation}\label{eq:arcs}
	A_{uv} = \begin{cases}
		\{(u,v)\} & \text{if} \ u=v_0 \\
		\{(v,u)\} & \text{if} \ v=v_0 \\
		\{(u,v),(v,u)\} & \text{otherwise}
	\end{cases}, \quad \quad \forall (u,v) \in E_0
\end{equation}
and
\begin{equation}
	A = \bigcup_{(u,v) \in E_0} A_{uv}\,.
\end{equation}

Let $T_D = (V_0,A_T)$ be a spanning arborescence of $G_D$, where $A_T \subsetneq A$.
For each arc $(u,v) \in A$, we define a binary variable $e_{uv}$ specifying if the arc is in $A_T$, thus following the definition for the same variable in the QUBO formulation for the Degree-Constrained Minimum Spanning Tree of \cite{fowler}.
In order to $T_D$ be a valid spanning arborescence, all of the following constraints must be met.

\subsection{Vertex constraints}\label{subsec:nodal_consts}
Each vertex in a spanning arborescence has exactly one incoming arc, with the exception of the root, as specified by the constraint
\begin{equation}\label{eq:nodal}
	\sum_{u \in N_{G_0}(v)} e_{uv} = 1, \quad \forall v \in V_0\setminus\{v_0\}
\end{equation}
where $N_{G_0}(v)$ is the set of all neighbours of $v$ in $G_0$.
The vertex constraints also follow the same definition as in \cite{fowler}.

\subsection{Edge constraints}
It is not possible to have in an arborescence both arcs of an undirected edge since this would close a 2-cycle.
This condition is prevented with the constraint
\begin{equation}\label{eq:edge_constr}
	e_{uv} e_{vu} = 0 \quad \forall (u,v) \in E_0, \; u,v \neq v_0\,.
\end{equation}

\subsection{Cycle constraints}
The previous two types of constraints impose that the number of arcs is equal to the number of vertices minus one.
While this condition is necessary to form a spanning arborescence, is not sufficient since the selected arcs may, with the previous constraints alone, still close directed cycles in $G_D$ and, equivalently, separate $T_D$ into a disconnected subgraph.
Thus, additional constraints are needed to prevent the closing of any possible cycle.

\begin{figure}
	\centering
	\begin{tikzpicture} [thick,scale=1.2]
		\graph [no placement] {
			{[y=1] 1[draw,x=0] -> 2[x=1]} -- 5[x=2] -- 7[x=4] -- {[y=1] 20[x=6] <- 0/1[draw,x=7]};
			2 -- {[y=2] 28[x=2] -- 14[x=3] -- 11[x=5]} -- 20;
			14 -- 8[x=4,y=1] -- 11;
			5 -- 28;
			8 -- 7;
		};
		\node at (1.6,1) {A};
		\node at (3,1) {B};
		\node at (4,1.6) {C};
		\node at (5,1) {D};
	\end{tikzpicture}
	\caption{The four cycles of the cycle basis of the reduced main biconnected component. The component root is vertex 1, shown twice to visually remove the facial cycle containing the root, which is excluded from the cycle basis. The root has outcoming arcs only. The remaining edges are undirected.}
	\label{fig:cycle_basis}
\end{figure}
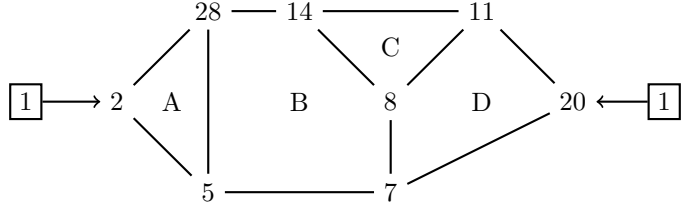

Since $G_0$ is a non-trivial biconnected component, each of its vertices and edges is in at least one cycle.
Assuming that $G_0$ is planar, as is typically the case of electrical distribution grids, the set of its facial cycles, excluding the outer face cycle, define a cycle basis in $G_0$.
Given that the root has no incoming arc in $G_D$, no directed cycle containing the root can be closed in $G_D$ since such a cycle would require at least one incoming arc (and one outcoming arc) incident to that vertex.
Thus, the facial cycles containing the root are excluded from the cycle basis considered for the cycle constraints.
Figure \ref{fig:cycle_basis} shows the four cycles (A, B, C and D) of the cycle basis of the reduced main biconnected component of the 33-node test network.

\subsubsection{Single cycle constraints}
There is a directed cycle in $G_D$ for each of the two possible directions of a cycle in $G_0$.
A constraint is needed to prevent the closure of each directed cycle, as specified by
\begin{subequations}\label{eq:single_cycle}
\begin{equation}
	\prod_{i=1 \dots n} e_{k_i k_{i+1}} = 0
\end{equation}
\begin{equation}
	\prod_{i=1 \dots n} e_{k_{i+1} k_i} = 0
\end{equation}
\end{subequations}
where $k_{n+1} \equiv k_1$ for a $n$-length cycle in $G_0$ given by the sequence of vertices $k_1,k_2,\dots,k_n,k_1 \in V_0$.

\subsubsection{Combined cycle constraints}
Let $C_1$ and $C_2$ be two cycles of the cycle basis sharing a contiguous path composed of one or more common edges and let $C_{12}$ be the combination of those two cycles.
$C_{12}$ is also a cycle and it is induced by the symmetric difference of the edge sets of $C_1$ and $C_2$, i.e., the union of these sets excluding the common edges.

Although the single cycle constraints applied for $C_1$ and $C_2$ prevent the closing of those cycles, this is not the case with $C_{12}$ since the arcs of the common edges of $C_1$ and $C_2$ would not be in $A_T$ and thus forcing the product of the above constraints to be zero for both $C_1$ and $C_2$.
As a consequence, $C_{12}$ would have to be considered as an additional cycle on those constraints in order to prevent its closure.

\begin{figure}
	\centering
	\begin{tikzpicture}[node distance=20mm, thick, main/.style = {draw, circle}]
		\node[main] (a) {A};
		\node[main] (b) [right of=a] {B};
		\node[main] (c) [above right of=b] {C};
		\node[main] (d) [below right of=c] {D};
		\draw (a) -- (b);
		\draw (b) -- (c);
		\draw (b) -- (d);
		\draw (c) -- (d);
	\end{tikzpicture}
	\caption{Graph representing the adjacency between the cycles of the cycle basis. An edge indicates two adjacent cycles.}
	\label{fig:cycles}
\end{figure}
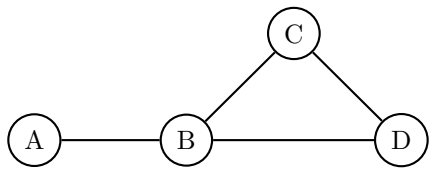

With multiple adjacent cycles, constraints must be added to each possible cycle combination.
Figure \ref{fig:cycles} shows the adjacency between the cycles of the cycle basis shown in figure \ref{fig:cycle_basis}. 
There are eight cycle combinations for this basis: AB, ABC, ABD, ABCD, BC, BD, CD and BCD.

\subsubsection{A new method for adjacent cycles}\label{sec:new_method}
As previously shown, the number of adjacent cycle combinations in our test network is twice the number of cycles in the cycle basis.
For larger bases, the number of combinations grows more than linearly with the basis size, resulting in an even larger number of additional cycle constraints being added to the model.

We are proposing a new method for adjacent cycle constraints without the need to consider the cycles formed by combinations of adjacent cycles.
This method is based on the idea of adding an auxiliary direction variable $d_{uv}$ to each edge $(u,v)$ common to two adjacent cycles $C_1$ and $C_2$.
This variable assigns a direction to the edge even if none of its two arcs is in $A_T$ (and thus defining a direction).
Since $(u,v)$ is an undirected edge, $d_{uv}$ is defined for the convention $u < v$. 
Its constraints are given by
\begin{subequations}\label{eq:def_d}
	\begin{align}
		e_{uv} &\implies d_{uv} \\
		e_{vu} &\implies \neg d_{uv}
	\end{align}
\end{subequations}
where $\neg d_{uv} \equiv 1 - d_{uv}$ is the logical negation of $d_{uv}$.
For this edge, these constraints replace the edge constraint (\ref{eq:edge_constr}) since they already prevent $e_{uv}$ and $e_{vu}$ from being both true.

From these constraints it is clear that if no arc of the edge $(u,v)$ is in $A_T$ then the value of $d_{uv}$ is not imposed by these constraints.
This is the case when $C_1$ and $C_2$ are not closed but $C_{12}$ is.
To prevent the closure of $C_{12}$ while still preventing $C_1$ and $C_2$ from closing, the single cycle constraints (\ref{eq:single_cycle}) for these two cycles must be changed by replacing $e_{uv}$ with $d_{uv}$ or $\neg d_{vu}$, as given by
\begin{subequations}\label{eq:adj_cycles}
	\begin{eqnarray}
		\prod_{i=1 \dots n} e'_{k_i k_{i+1}} = 0 \\
		\prod_{i=1 \dots n} e'_{k_{i+1} k_i} = 0
	\end{eqnarray}
\end{subequations}
\begin{equation*}
	\text{where} \ e'_{uv} \equiv \left\{\begin{array}{rl}
		d_{uv} & \text{if} \  u < v \wedge (u,v) \in E(C_1) \cap E(C_2)\\
		\neg d_{vu} & \text{if} \  v < u \wedge (u,v) \in E(C_1) \cap E(C_2)\\
		e_{uv} & \text{otherwise}\end{array}\right..
\end{equation*}

\begin{figure}
	\centering
	\begin{tikzpicture} [thick,scale=1.5]
		\graph [no placement] {
			1[x=0,y=1] -> 2[x=1,y=2] -> 3[x=2,y=1] -> 4[x=1,y=0] -> 1;
			2 <->[dashed] 4;
		};
		\node at (0.6,1) {$C_1$};
		\node at (1.4,1) {$C_2$};
	\end{tikzpicture}
	\caption{$C_{12}$ is closed (in the clockwise direction). Since none of the arcs (2,4) and (4,2) is in $A_T$, the direction variable $d_{24}$ can assume any value. If $d_{24}=1$, the $C_1$ constraint is violated; otherwise, the $C_2$ constraint is violated.}
	\label{fig:cycle_combination}
\end{figure}
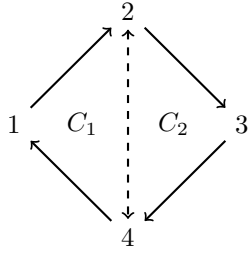

Given the logical implications of (\ref{eq:def_d}), it is clear that (\ref{eq:adj_cycles}) still prevents the closure of $C_1$ and $C_2$.
The closure of $C_{12}$ is also prevented since one of the constraints for either $C_1$ or $C_2$ would be violated, depending on the value of $d_{uv}$ which, in this situation, is not imposed by (\ref{eq:def_d}).
In other words, (\ref{eq:adj_cycles}) together with (\ref{eq:def_d}) are stricter than (\ref{eq:single_cycle}) alone because every common edge must have a direction regardless of having any of its arcs in $A_T$ or not, and that direction will violate one of the cycle constraints if $C_{12}$ is closed, as illustrated in figure \ref{fig:cycle_combination}.

\subsubsection{Generalization to cycles with interior vertices}
While an edge cannot be in more than two facial cycles, a vertex can be shared by more than two of such cycles.
This vertex may lie in the outer face cycle (which is not part of the cycle basis), or
otherwise be an interior vertex.
As mentioned in \ref{sec:edge_lifting}, all graph vertices after edge lifting have a degree of at least 3, excluding the root vertex.
Then, any interior vertex is necessarily in three or more cycles.
In our test network cycle basis, shown in figure \ref{fig:cycle_basis}, vertex 8 is the only interior vertex and thus it is the only vertex common to at least three cycles (namely, cycles B, C and D).

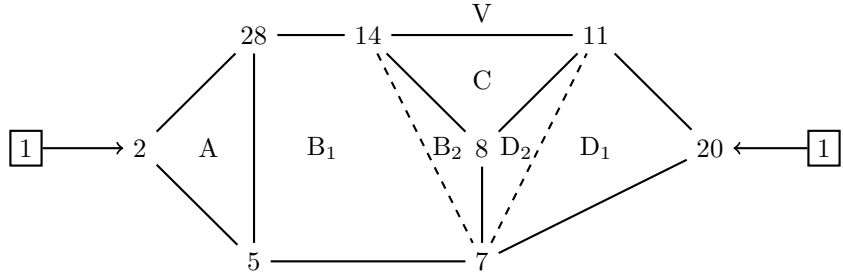
\begin{figure}
	\centering
	\begin{tikzpicture} [thick,scale=1.5]
		\graph [no placement] {
			{[y=1] 1[draw,x=0] -> 2[x=1]} -- 5[x=2] -- 7[x=4] -- {[y=1] 20[x=6] <- 0/1[draw,x=7]};
			2 -- {[y=2] 28[x=2] -- 14[x=3] -- 11[x=5]} -- 20;
			14 -- 8[x=4,y=1] -- 11;
			5 -- 28;
			8 -- 7;
			14 --[dashed] 7 --[dashed] 11;
		};
		\node at (1.6,1) {A};
		\node at (2.6,1) {B$_1$};
		\node at (4,1.6) {C};
		\node at (5,1) {D$_1$};
		\node at (4,2.2) {V};
		\node at (3.7,1) {B$_2$};
		\node at (4.3,1) {D$_2$};
	\end{tikzpicture}
	\caption{The cycle basis with the virtual cycle V and the new inner cycles B$_2$ and D$_2$. Dashed lines represent the virtual cycle edges not in $E_0$.}
	\label{fig:virtual_cycle}
\end{figure}

The closure of a combination of cycles containing all cycles sharing an interior vertex -- as is the case of combinations BCD and ABCD in our test network -- is not prevented by the constraints defined in \ref{sec:new_method} alone.
To handle this case, a virtual cycle must be added such that it encloses the interior vertex.
The cycle vertices are the neighbors of the interior vertex and the cycle edges may or may not be in $E_0$.
If any of these neighbors is also an interior vertex, a new virtual cycle is added to enclose this vertex, and the procedure is repeated until there are no further interior vertices to enclose.
To illustrate this method with the test network, a virtual cycle is added to enclose the interior vertex 8, as shown in figure \ref{fig:virtual_cycle}.

The new virtual cycle V is defined by the neighbors of vertex 8 -- vertices 7, 11 and 14 -- and the following relations hold:
\begin{equation}
	\begin{split}
		\text{V} & = \text{C} \oplus \text{B}_2 \oplus \text{D}_2 \\
		\text{B} & = \text{B}_1 \oplus \text{B}_2 \\
		\text{D} & = \text{D}_1 \oplus \text{D}_2
	\end{split}
\end{equation}
where $\oplus$ represents the combination of two cycles.
From the three edges of cycle V, only the edge (11,14) is in $E_0$.
Since this edge is common to two cycles (V and C), a $d_{11,14}$ variable is allocated as previously described.
The other two edges, represented by the dashed lines in figure \ref{fig:virtual_cycle}, are shared by three cycles each -- edge (7,14) is common to cycles V, B$_1$ and B$_2$, while edge (7,11) is common to V, D$_1$ and D$_2$.
Variables $d_{7,14}$ and $d_{7,11}$ are also allocated for these two edges, but since these edges are not in $E_0$, no $e$ variables are allocated to them and thus the constraints defined in (\ref{eq:def_d}) do not apply to these edges.

All cycles in figure \ref{fig:virtual_cycle} except V are facial cycles, and thus they compose a valid cycle basis.
Applying the constraints defined in (\ref{eq:adj_cycles}) to these facial cycles alone would yield the same results as with the original cycle basis, where the closure of (A)BCD is not prevented.
On the other side, cycle V may be viewed as a shorter representation of (A)BCD since those constraints force the closure of the former if and only if the latter is closed.
Then, applying these constraints to cycle V as well will finally prevent the closure of (A)BCD.

\subsection{Path constraints}
The edge lifting reduction described in section \ref{sec:edge_lifting} assigns a path $P_{uv} \subset G_C$ for each edge $(u,v)$ of the topological minor $G_0$ of $G_C$, such that a network configuration in $G_C$ can be represented by the pair $(T_0,Q)$, where $T_0$ is a spanning tree of $G_0$ and $Q$ is the set of open edges of $G_C$ (i.e., the edges not in the spanning tree $T_C$ equivalent to $(T_0,Q)$).
A network configuration can be equally represented by the spanning arborescence $T_D$ instead of $T_0$ in the pair $(T_D,Q)$

In order to indirectly specify the set $Q$, a binary variable $p_x$ is added for each inner vertex $x$ of a path $P_{uv}$ (thus excluding the end vertices $u$ and $v$), for all edges of $G_0$.
Since $P_{uv}$ is undirected, the convention $u < v$ is assumed in the following discussion.
Let the path $P_{uv}$ with $n$ inner vertices be represented by the vertex sequence $u,k_1,k_2,\dots,k_n,v$.
The $p_{k_i}$ variable specifies which of the end vertices -- $u$ (if 1) or $v$ (if 0) -- is upward of the inner vertex $k_i$, with $i=1\dots n$.

If the path is open, let $k_i$ and $k_{i+1}$ be the end vertices of the open edge within the path.
Then, the inner vertices $k_1,\dots,k_i$ have $u$ as the upward end vertex, and the inner vertices $k_{i+1},\dots,k_n$ have $v$ as the upward end vertex.
Thus, the open edge is the only one in the path where $(p_{k_i}, p_{k_{i+1}}) = (1,0)$, $\exists i=1,\dots,n-1$.
Also, the condition $(p_{k_i}, p_{k_{i+1}}) = (0,1)$, $\exists i=1,\dots,n-1$ is never allowed in any path, as specified for all paths $P_{uv}$ with $n^{uv} > 1$ inner nodes by the constraint
\begin{equation}\label{eq:path}
	p_{k^{uv}_{i+1}} \implies p_{k^{uv}_i}, \quad \forall (u,v) \in E_0, u < v, i=1,\dots,n^{uv}-1
\end{equation}
where $k^{uv}_i$ is the $i$-th inner node of path $P_{uv}$ counted from $u$ to $v$.

If a path $P_{uv}$ is closed, either $e_{uv}$ or $e_{vu}$ specifies the path direction and all the path inner vertices have the same upward end vertex, either $u$ or $v$ respectively, as specified in (\ref{eq:path}) together with the constraints
\begin{subequations}\label{eq:path1}
	\begin{align}
		e_{uv} &\implies p_{k^{uv}_{n^{uv}}}\\
		e_{vu} &\implies \neg p_{k^{uv}_1}
	\end{align}
\end{subequations}
defined for any path $P_{uv}$ with $n^{uv} > 0$ inner nodes.

\subsection{Auxiliary load-arc variables}\label{subsec:aux_vars}
For the sake of clarity in the discussion, this section first explains the method for the auxiliary load-arc variables assuming that $G_0 \equiv G_C$, i.e., that no edge lifting was performed to transform $G_C$ into $G_0$.
Later, the edge lifting transformation is then considered.

\subsubsection{Formulation without edge lifting}
As formulated in (\ref{eq:loads}), the set $D_{uv}(T,v_0)$ (of all vertices downward of a link $(u,v) \in E(T)$ across the spanning tree $T$) is needed in order to compute the electrical current flowing through the link.
Given that the problem constraints are formulated in terms of a spanning arborescence $T_D = (V_0,A_T)$, we consider instead the set $D_{uv}(T_D,v_0)$ of all vertices downward of an arc $(u,v) \in A_T$ across $T_D$.
An auxiliary binary variable $z_{uvn}$ is defined to specify whether a vertex $n$ is in $D_{uv}(T_D,v_0)$, i.e., whether the load of $n$ contributes to the current on the arc $(u,v)$ (thus giving the name load-arc variables).
These auxiliary variables are needed to express the network power losses in the QUBO model, as will be described in the next section.

Since $A_T$ and thus $T_D$ are defined by the $e_{uv}$ variables previously introduced, each $z_{uvn}$ variable is ultimately a function of $e$ variables.
Assigning a $z_{uvn}$ variable from $e$ variables alone would lead to extremely complex boolean expressions since all possible paths in $G_D$ between vertices $v$ and $n$ would need to be explicitly accounted for.
This would defeat the whole purpose of the QUBO optimization since the complete search space would need to be traversed just to build the QUBO model.

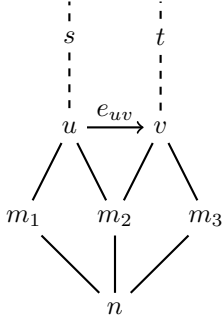
\begin{figure}
	\centering
	\begin{tikzpicture}[scale=1.2, thick]
		\node (a) at (-.5,1.5) {};
		\node (b) at (.5,1.5) {};
		\node (s) at (-.5,1) {$s$};
		\node (t) at (.5,1) {$t$};
		\node (u) at (-.5,0) {$u$};
		\node (v) at (.5,0) {$v$};
		\node (m1) at (-1,-1) {$m_1$};
		\node (m2) at (0,-1) {$m_2$};
		\node (m3) at (1,-1) {$m_3$};
		\node (n) at (0, -2) {$n$};

		\draw[dashed] (s) -- (a);
		\draw[dashed] (t) -- (b);
		\draw[dashed] (s) -- (u);
		\draw[dashed] (t) -- (v);
		\draw[->] (u) -- node[midway, yshift=2mm] {$e_{uv}$} (v);
		\draw (u) -- (m1);
		\draw (u) -- (m2);
		\draw (v) -- (m2);
		\draw (v) -- (m3);
		\draw (n) -- (m1);
		\draw (n) -- (m2);
		\draw (n) -- (m3);		
	\end{tikzpicture}
	\caption{Example of the relationship between vertex $n$ and arc $(u,v)$.}
\label{fig:aux_vars}
\end{figure}

We are proposing a novel method to assign the $z$ variables with minimal overhead.
The central concept of this method is having these variables depending not only on $e$ variables but on other neighboring $z$ variables as well.
As an illustration of this method, vertex $n$ in figure \ref{fig:aux_vars} contributes to the current in arc $(u,v)$ (as specified by $z_{uvn}$) if and only if the arc is in $T_D$ (as stated by $e_{uv}$) and if the vertex also contributes to the current in arcs $(v,m_2)$ or $(v,m_3)$.
Since no cycles are allowed in the arborescence, $n$ contributes to at most one of these two arcs.
The complete definition is then $z_{uvn} = e_{uv}(z_{vm_2n} + z_{vm_3n})$.
Similarly, for the arc in the opposite direction, $z_{vun} = e_{vu}(z_{um_1n} + z_{um_2n})$.

The $z$ variables are then more generally defined as
\begin{equation}\label{eq:aux_vars}
	z_{uvn} = e_{uv} \sum_{m \in N_{G_0}(v) \setminus \{u,v_0\}} z_{vmn} \quad \forall (u,v) \in A, n \in V_0 \setminus \{u,v,v_0\}\,.
\end{equation}

One may notice that this formulation applied to the given example includes $z_{vtn}$ and $z_{usn}$ in the definition of $z_{uvn}$ and of $z_{vun}$, respectively, but it is easy to realize that both $z_{vtn}$ and $z_{usn}$ are always zero.
This formulation does not define $z_{uvn}$ if $n = v$, but in the expression $z_{vmn}$, $m$ may be equal to $n$.
Thus, to define a $z_{xyy}$ variable, one may realize that it is always equal to $e_{xy}$ and in fact they are one and the same variable in the QUBO model, as stated by
\begin{equation}\label{eq:zuvv}
	z_{uvv} \equiv e_{uv}, \quad \forall (u,v) \in A\,.
\end{equation}

\subsubsection{Formulation with edge lifting}\label{sec:zvars_edgelifting}
As defined in (\ref{eq:aux_vars}), a $z_{uvn}$ variable is allocated for each combination of an arc in $A$ with all vertices of $V_0$ excluding the arc end-vertices and $v_0$, thus the number of such variables is $|A|\left(|V_0|-3\right)+d_{G_0}(v_0)$, where $d_{G_0}(v_0)$ is the degree of $v_0$ in $G_0$.
The application of (\ref{eq:aux_vars}) and (\ref{eq:zuvv}) to the main biconnected component of our test network without edge lifting would result in 2032 $z$ variables.
Given the size limits of current quantum annealers, it is crucial to reduce the number of these variables.
This is the main motivation to use edge lifting to reduce $G_C$ into the topological minor $G_0$.

Since an arc $(u,v) \in A$ represents a path $P_{uv} \subset G_C$ where each link in the path has a distinct current value, in general there is no single current value on the arc (unless the path has one link only).
Thus, with edge lifting, a $z_{uvn}$ variable shall be interpreted as whether a vertex $n \in V_C$ contributes to the current of the links in path $P_{uv}$ along the direction defined by arc $(u,v)$.

Given that $V_0 \subset V_C$, (\ref{eq:aux_vars}) and (\ref{eq:zuvv}) applied to $G_D$ would not define a $z_{uvn}$ variable (and thus a $z_{vmn}$ variable) when vertex $n \in V_C$ and $n \notin V_0$, although its load current would also contribute to the current of some network links.
Such a vertex is necessarily an inner vertex of some path $P_{uv}$ where $(u,v) \in E_0$.
In this case, just like a $z_{uvv}$ variable, $z_{uvn}$ does not exist as an explicit QUBO variable.
Instead, it is defined as
\begin{equation}\label{eq:zuvn_path}
	z_{uvn} \equiv 
	\left\{\begin{array}{rl}
		p_n & \text{if} \ u < v \\
		\neg p_n & \text{if} \ u > v\end{array}\right.
	\quad \forall (u,v) \in \bigcup_{(a,b) \in E_0} \{(a,b),(b,a)\},\ n \in V(P_{uv}) \setminus \{u,v\} \,.
\end{equation}

This definition applies to both directions of every edge in $E_0$, instead of being applied to all arcs in $A$, since $z_{uvn}$ variables are used in the power losses function and this function applies to both directions of every path $P_{uv}$ including the ones incident to the root vertex, as will be seen in \ref{sec:power_losses}.

Finally, (\ref{eq:aux_vars}) must be generalized to define $z_{uvn}$ when vertex $n \in V_C, n \notin V_0$ is an inner vertex of a path other than $P_{uv}$, as given by
\begin{equation}\label{eq:aux_vars1}
	z_{uvn} = e_{uv} \sum_{m \in N_{G_0}(v) \setminus \{u,v_0\}} z_{vmn} \quad \forall (u,v) \in A, n \in V_C \setminus \left(V(P_{uv}) \cup \{v_0\}\right)\,.
\end{equation}

With (\ref{eq:zuvv}), (\ref{eq:zuvn_path}) and (\ref{eq:aux_vars1}), $z_{uvn}$ is now defined for any arc $(u,v) \in A$ and for any vertex $n \in V_C \setminus \{u,v_0\}$. The hypothetical $z_{uvu}$ and $z_{uvv_0}$ variables would always be zero and thus they do not need to be defined.
With edge lifting, the number of $z$ variables for the main biconnected component of our test network reduces to 654.

A $z_{uvn}$ variable is always zero when, due to the network topology, there is no spanning arborescence such that vertex $n$ contributes to the current in arc $(u,v)$.
Such $z$ variables can be excluded from the model, resulting in a further reduction of the number of $z$ variables to 577.

\section{QUBO Formulation}
\label{section:formulation}
A QUBO model formulates a pseudo-Boolean function $f:\mathbb{B}^n \rightarrow \mathbb{R}$ \cite{BOROS2002155} as a quadratic polynomial over $n$ binary variables $x_i\in \mathbb{B}$, where $\mathbb{B}=\{0,1\}$:
\begin{equation}\label{eq:qubo}
f(x)=\sum_{i=1}^n a_i x_i + \sum_{i<j}b_{ij} x_i x_j + c	
\end{equation}
where $a_i$ and $b_{ij}$ are the real-valued linear and quadratic coefficients, respectively, and $c$ is a constant term.
Solving a QUBO problem consists on finding the binary string $x^*$ which minimizes $f$:
\begin{equation}
x^* = \operatorname*{arg\,min}_x f(x)\,.
\end{equation}

While the constant term $c$ has no influence on $x^*$, it's inclusion on $f$ makes this function more general in order to provide meaningful cost values for the optimization problem at stake.

For our optimization problem, the solution $x^*$ contains the variables described in the previous section.
Thus, each of these variables is assigned to a given variable index $i$ in the QUBO model.

\subsection{Problem constraints}\label{sec:prob_constrs}
Since, by definition, the QUBO model is unconstrained, the problem constraints described in the previous section must be added to the model as penalty expressions.
To this end, these constraints are first written as a Constraint Satisfaction Problem (CSP) using the \texttt{dwavebinarycsp} Python package from the D-Wave Ocean SDK \cite{dwbinarycsp}.
The CSP is then converted to the QUBO model using the \texttt{stitch} function from this package with the default minimum penalty value of 2.0.
Thus, the violation of any constraint would increase the value of the QUBO model by at least 2.0, while all feasible solutions keep the QUBO value unchanged. 
The QUBO model is represented by a \texttt{BinaryQuadraticModel} class instance on which the power losses terms are added.

\subsection{Power losses function}\label{sec:power_losses}
Network power losses minimization is the goal of this constrained optimization problem.
As defined in \ref{sec:edge_lifting}, let $G_0 = (V_0,E_0)$ be a topological minor of a non-trivial biconnected component $G_C=(V_C,E_C)$ of the original network $G$.
A power losses function can be defined for each edge $(u,v) \in E_0$ as the sum of the losses in all links $(a,b)$ of the path $P_{uv} \subset G_C$.
This function depends on the network configuration, which can be represented by the spanning tree $T_C$ of $G_C$, as given by
\begin{equation}\label{eq:losses0}
	L^0_{uv}(T_C) = \sum_{(a,b) \in E(P_{uv})} L_{ab}(T_C) \quad \forall (u,v) \in E_0
\end{equation}
where the losses function $L_{ab}(T_C)$ is defined as in (\ref{eq:losses}).
Given that each link $(a,b) \in E_C$ is in exactly one path $P_{uv}$, the total network losses $\sum_{(a,b) \in E_C} L_{ab}$ are same as $\sum_{(u,v) \in E_0} L^0_{uv}$ and thus the original optimization problem defined in (\ref{eq:optim}) for the component $G_C$ can be reformulated in terms of $L^0_{uv}$.

With the auxiliary load-arc variables introduced in the previous section, the power losses function can be expressed as a sum of linear and quadratic terms of these variables and thus these terms can be directly added to the QUBO model.
As formulated in (\ref{eq:losses}), the power losses in a given link are quadratic with the current flowing on the link and, as stated in (\ref{eq:loads}), this current is the sum of the currents from all loads downward of the link.
Thus, the expansion of the square of the sum yields quadratic terms for all pairwise combinations between the sum terms.

The losses $L^0_{uv}$ can defined as the sum of two directed losses
\begin{equation}
	L^0_{uv} = L^D_{uv} + L^D_{vu} \quad \forall (u,v) \in E_0
\end{equation}
where $L^D_{uv}$ represents the total losses on the edges of $P_{uv}$ downward of vertex $u$ and $L^D_{vu}$ corresponds to the losses on the edges downward of $v$.
If the path is closed then all the links in the path are downward of either $u$ or $v$ and then one of the directed losses is zero.
The directed losses in an arc $(u,v)$ of each of the two directions of an edge in $E_0$ can be expressed in terms of QUBO variables, as given by
\begin{equation}\label{eq:losses_qubo}
\begin{split}
L^D_{uv} = & e_{uv} \left( R_{uv} \left|I^L_v\right|^2 + L'_{uv}\left(I^L_v\right) \right) \\
& + \quad \sum_{n \in V(P_{uv}) \setminus \{u,v\}} z_{uvn} \left( R_{un} \left|I^L_n\right|^2 + \sum_{k \in V(P_{uv}n) \setminus \{u,n\}} 2 R_{uk} \Re\left(I^L_n \overline{I^L_k} \right) \right) \\
& + \sum_{n \in V_C \setminus (V(P_{uv}) \cup \{v_0\})} z_{uvn} \left[ R_{uv} \left(\left|I^L_n\right|^2 + 2\,\Re\left(I^L_n \overline{I^L_v} \right) \right) + L'_{uv}\left(I^L_n\right) \right] \\
& + \sum_{\substack{n,k \in V_C \setminus (V(P_{uv}) \cup \{v_0\}) \\ n \neq k}} z_{uvn} z_{uvk}\,2 R_{uv} \Re\left( I^L_n \overline{I^L_k} \right)
\end{split}
\end{equation}
where
\begin{align*}
	L'_{uv}\left(I\right) & \equiv \sum_{k \in V(P_{uv}) \setminus \{u,v\}} 2 R_{uk} \Re\left(I \overline{I^L_k} \right) \\
	R_{ux} & \equiv \sum_{(a,b) \in E(P_{uv}x)} R_{ab}
\end{align*}
with $\Re(c)$ and $\overline{c}$ being the real part and the complex conjugate, respectively, of a complex number $c$, and with $P_{uv}x \subseteq P_{uv}$ being the path along $P_{uv}$ between vertices $u$ and $x$ where $x \in V(P_{uv}) \setminus \{u\}$.

Given the set $V(P_{uv}) \setminus \{u,v\}$ for the vertex $n$ in the first sum of the second line of (\ref{eq:losses_qubo}), the $z_{uvn}$ variables in this line are defined in (\ref{eq:zuvn_path}) as being equivalent to either $p_n$ or $\neg p_n$, while the remaining $z$ variables in (\ref{eq:losses_qubo}) are defined in (\ref{eq:aux_vars1}) as explicit variables.
Note that $L^D_{uv}$ is defined for any arc $(u,v) \in \bigcup_{(a,b) \in E_0} \{(a,b),(b,a)\} \supsetneq A$.
If $v=v_0$ in such an arc, then $(u,v) \notin A$ and all $e$ and explicit $z$ variables in (\ref{eq:losses_qubo}) are not defined and thus they are considered to be zero.
As a consequence, only the terms in the second line of (\ref{eq:losses_qubo}) are nonzero in this situation.

Before being added to the QUBO model, the losses terms are scaled by a suitable constant factor.
Applying a constant scaling to the network losses does not change the optimal solution of the original problem defined in (\ref{eq:optim}).
The purpose of this scaling is to keep the optimal losses value below the minimum constraints penalty value of 2.0 (mentioned in Section \ref{sec:prob_constrs}), while maximizing the magnitude of the losses function to gain resolution on the quantum annealer since it has a limited dynamic range for the physical realization of the model coefficients \cite{boothby2021architectural}.
If the scaled optimal losses value were greater than the minimum constraints penalty value, the solution with minimum QUBO value could be infeasible since a constraint could be violated in order to unfeasibly decrease the network losses.

\section{Results}
\label{section:results}
In this section the results of the application of our formulation are presented and discussed.
These results include the QUBO model metrics for the 33-node test network and the validation of this model through the analysis of its optimal solution as found by classical solvers.

\subsection{QUBO model metrics}
\label{subsec:model_metrics}
The QUBO model for the 33-node test network has a total of 1074 variables, as detailed in table \ref{tab:vars}.
Variables $e$, $d$, $z$ and $p$ were already described in section \ref{sec:constraints}.
The $y$ variables were manually introduced to improve the conversion to QUBO of some $z$ variables constraints defined in (\ref{eq:aux_vars1}) by reducing the number of interactions between variables.
Finally, the auxiliary variables are automatically added in the conversion to QUBO of the cycle constraints defined in (\ref{eq:adj_cycles}) which involve the product of three or more variables.

\begin{table}
	\centering
	\begin{tabular}{c c}
		\hline
		\textbf{Variable class} & \textbf{Number of variables}\\
		\hline
		$e$ & 24 \\
		$d$ & 4 \\
		$p$ & 23 \\
		$z$ & 577 \\
		$y$ & 434 \\
		auxiliary & 12 \\
		\hline
		\textbf{Total} & \textbf{1074}\\
		\hline
	\end{tabular}
	\caption{QUBO model variable allocation per variable class.}
	\label{tab:vars}
\end{table}

The model has a total of 10166 interactions between variables (i.e., product terms between two variables), as detailed in table \ref{tab:inters}.
As formulated in (\ref{eq:losses_qubo}), the $z$ variables are needed for the network losses terms.
Together with their intermediary $y$ variables, they represent 94.1\% of the model variables, while their interactions within constraints (\ref{eq:aux_vars1}) and network losses (\ref{eq:losses_qubo}) account for 98.6\% of the model interactions.
The dominance of these variables in the model size justifies the effort in reducing the number of $z$ variables.
As described in \ref{sec:zvars_edgelifting}, this number was reduced from 2032 to 654 thanks to edge lifting and further reduced to 577 through elimination of null $z$ variables.

\begin{table}
	\centering
	\begin{tabular}{c c}
		\hline
		\textbf{Constraints type / Function} & \textbf{Interactions}\\
		\hline
		Vertex (\ref{eq:nodal}) & 24\\
		Edge (\ref{eq:edge_constr}) or (\ref{eq:def_d}) & 13\\
		Cycle (\ref{eq:adj_cycles}) & 72\\
		Path (\ref{eq:path}) & 14\\
		Edge-path (\ref{eq:path1}) & 17\\
		$z$ and $y$ variables (\ref{eq:aux_vars1}) & 3029\\
		Network losses (\ref{eq:losses_qubo}) & 6997\\
		\hline
		\textbf{Total} & \textbf{10166}\\
		\hline
	\end{tabular}
	\caption{QUBO model interactions count per constraints type or per function.}
	\label{tab:inters}
\end{table}

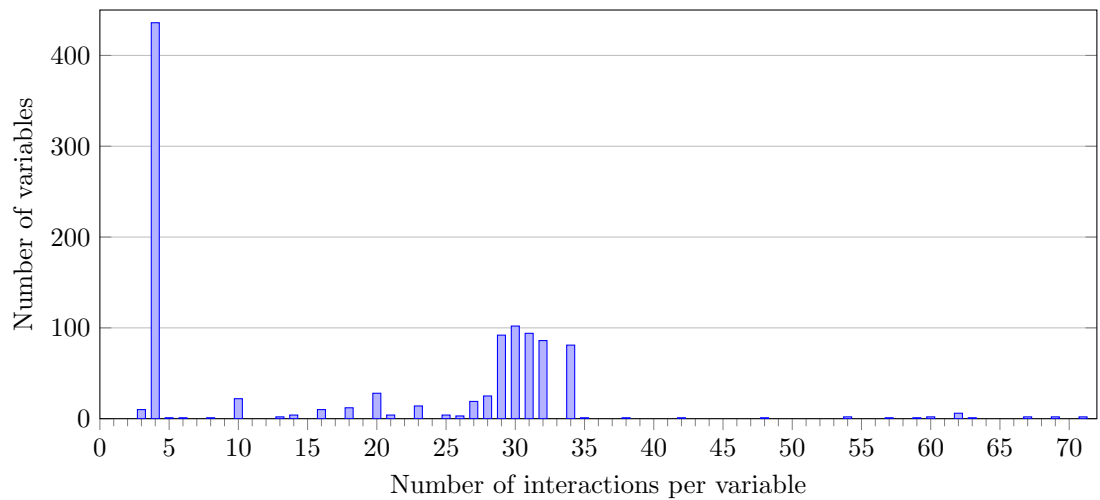
\begin{figure}
	\centering
	\begin{tikzpicture}
	\begin{axis}[
		width=\textwidth,
		height=7cm,
		xmin=0,xmax=72,ymin=0,ymax=450,
		xlabel={Number of interactions per variable},
		ylabel={Number of variables},
		ymajorgrids,
		xtick pos=left,
		minor x tick num=4,
		ybar,
		bar width=3pt,
		]
		\addplot
		coordinates {
		(3, 10)
		(4, 436)
		(5, 1)
		(6, 1)
		(8, 1)
		(10, 22)
		(13, 2)
		(14, 4)
		(16, 10)
		(18, 12)
		(20, 28)
		(21, 4)
		(23, 14)
		(25, 4)
		(26, 3)
		(27, 19)
		(28, 25)
		(29, 92)
		(30, 102)
		(31, 94)
		(32, 86)
		(34, 81)
		(35, 1)
		(38, 1)
		(42, 1)
		(48, 1)
		(54, 2)
		(57, 1)
		(59, 1)
		(60, 2)
		(62, 6)
		(63, 1)
		(67, 2)
		(69, 2)
		(71, 2)};
	\end{axis}
	\end{tikzpicture}
	\caption{Distribution of the number of interactions per model variable}
	\label{fig:histogram}	
\end{figure}

Figure \ref{fig:histogram} shows the distribution of the number of interactions per variable.
This metric is important for the embedding of the QUBO model in a quantum annealer \cite{cai2014practical}.
This distribution spans from 3 to 71 interactions per variable, with an average of 18.9 and a sharp peak of 436 variables with 4 interactions each.
Considering the variable classes, this distribution is partitioned into relatively well defined regions, as shown in table \ref{tab:var_niters}.

\begin{table}
	\centering
	\begin{tabular}{c c}
		\hline
		\textbf{Variable class} & \textbf{Interactions per variable (number of variables)}\\
		\hline
		auxiliary & 3(10), 4(2) \\
		$y$ & 4(434) \\
		$p$ & 5(1), 6(1), 10(21) \\
		$d$ & 8(1),  10(1), 13(2)\\
		$z$ & 14\dots34(577) \\
		$e$ & 31(1), 35\dots71(23)\\
		\hline
	\end{tabular}
	\caption{Distribution of the number of interactions per variable for each variable class.}
	\label{tab:var_niters}
\end{table}

\subsubsection{Comparison with other formulations}
Our model cannot be completely compared with other QUBO formulations found in the literature since no other known formulation includes an optimization objective with a quadratic cost over the network flows.
Still, the part of the model relative to the constraints ensuring a valid spanning tree solution can be compared with the equivalent function of other formulations.
Tables \ref{tab:nvars} and \ref{tab:niters} compare the number of variables and the number of interactions, respectively, needed for such constraints between spanning tree problem formulations found in the literature and our model.
This comparison considers the non-trivial biconnected component $G_C$ of the 33-node test network and its reduced form $G_0$ obtained from the edge lifting procedure (figure \ref{fig:comp_reduced}), where $G_C=(V_C,E_C)$ and $G_0=(V_0,E_0)$ with $|V_C| = 32$, $|E_C| = 36$, $|V_0| = 9$ and $|E_0| = 13$.
The metrics with $G_0$ enable the comparison between formulations without taking into account the effect of edge lifting (i.e., considering only vertex, edge and cycle constraints), while the metrics with $G_C$ include this effect leveraging on the existence of linear chains which are typically found in electrical networks.
Both with $G_0$ and with $G_C$, our model allocates less variables than the other formulations and it shows a linear scaling with the network size, while the other formulations have a quadratic scaling (assuming $\mathcal{O}(|E|)$ equivalent to $\mathcal{O}(|V|)$).
Our model also allocates less interactions for both graphs with a linear scaling while the other formulation shows a cubic scaling.
The advantage with $G_C$ is particularly striking given the additional effect of edge lifting on the linear chains of this graph.

\begin{table}
	\centering
	\begin{tabular}{l c @{}c c c}
		\hline
		\textbf{Model} & \textbf{Number of variables} & \textbf{With $G_0$} & \textbf{With $G_C$} & \textbf{Scaling}\\
		\hline
		\cite{Lucas_2014}\footnotemark & $|V|\lfloor(|V|+3)/2\rfloor + |E|(|V|+1)$ & 184 & 1732 & $\mathcal{O}(|V||E|)$\\
		\cite{fowler} & $2|E| - |N_G(v_0)| + \binom{|V|-1}{2}$ & 52 & 535 & $\mathcal{O}(|V|^2)$\\
		\multirow{2}{*}{Ours} & $|e| + |d| + |\text{auxiliary}|$ & 40 & -- & $\mathcal{O}(|E|)$\\
		& $|e| + |d| + |p| + |\text{auxiliary}|$ & -- & 63 & $\mathcal{O}(|E|)$\\
		\hline
	\end{tabular}
	\caption{Number of variables needed for the (degree-constrained) minimum spanning tree formulations in \cite{Lucas_2014} and in \cite{fowler}, and number of variables allocated in our model for spanning tree constraints in $G_0$ and $G_C$.
	To enable a fair comparison, the terms of the degree constraints were removed from the first two formulations since our model does not include such constraints.
	Given that \cite{fowler} and our model have the same definition for the $e$ variables, the term $2|E| - |N_G(v_0)|$ is equal to $|e|$.
	The number of auxiliary variables in our model grows linearly with the total number of variables in cycle constraints (i.e., total number of arcs in basis cycles) in excess of three variables per constraint, thus a linear scaling with $|E|$ is assumed for $|\text{auxiliary}|$ as for $|e|$, $|d|$ and $|p|$.
	}
	\label{tab:nvars}
\end{table}

\footnotetext{The original expression in \cite{Lucas_2014} has a minor mistake. The correct expression for the formulation of \cite{Lucas_2014} shown here is actually given in \cite{fowler} while citing that formulation.}

\begin{table}
	\centering
	\begin{tabular}{l c @{}c c c}
		\hline
		\textbf{Model} & \textbf{Number of interactions} & \textbf{With $G_0$} & \textbf{With $G_C$} & \textbf{Scaling}\\
		\hline
		\cite{fowler} & $\sum_{v\in V\setminus \{v_0\}} \binom{|N_G(v)|}{2} + 2(|E| - |N_G(v_0)|) + 3\binom{|V|-1}{3}$ & 214 & 13600 & $\mathcal{O}(|V|^3)$\\
		\multirow{2}{*}{Ours} & NI(vertex con.) + NI(edge con.) + NI(cycle con.) & 109 & -- & $\mathcal{O}(|E|)$\\
		& same as above + NI(path c.) + NI(edge-path c.) & -- & 140 & $\mathcal{O}(|E|)$\\
		\hline
	\end{tabular}
	\caption{Number of interactions needed for spanning tree constraints in $G_0$ and $G_C$ for the formulation of \cite{fowler} and for our model.
	No number of interactions was provided in \cite{Lucas_2014}.
	As in table \ref{tab:nvars}, the terms of the degree constraints were removed from the first formulation.
	NI represents the number of interactions for the given constraints class in our model (table \ref{tab:inters}).
	Since both models have the same definition for vertex constraints, the term $\sum_{v\in V\setminus \{v_0\}} \binom{|N_G(v)|}{2}$ is equal to NI(vertex con.).
	Given than the average vertex degree is not expected to grow with larger networks, $\mathcal{O}(|E|)$ and $\mathcal{O}(|V|)$ are considered equivalent and a linear scaling with $|E|$ is assumed for NI(vertex con.) as for NI(edge con.), NI(cycle con.), NI(path con.) and NI(edge-path con.).  
	}
	\label{tab:niters}
\end{table}

\subsubsection{Scaling of the complete model}
The complete QUBO model includes the network losses terms which, as described in \ref{subsec:model_metrics}, dominate the model in terms of both the number of variables and the number of interactions.
The $z$ variables -- and their intermediary $y$ variables -- are needed to compute the network losses as a QUBO value.
As mentioned in \ref{sec:zvars_edgelifting}, a variable $z_{uvn}$ is defined for every pair of an arc $(u,v) \in A$ with a vertex $n \in V_C \setminus \{u,v_0\}$.
As defined in \ref{sec:constraints}, $A$ is the set of arcs in $G_D$ (the directed graph associated with $G_0$).
Given that $|A| \in \mathcal{O}(|E_0|)$, the number of $z$ variables is $\mathcal{O}(|V_C| |E_0|)$.
The number of $y$ variables has the same order.

The product terms of the network losses, as defined in the last line of (\ref{eq:losses_qubo}), account for the great majority of the model interactions, as seen in table \ref{tab:inters}.
For every arc $(u,v)$, each pair of distinct vertices $n,k \in V_C \setminus \left(V(P_{uv}) \cup \{v_0\}\right)$ originates a product $z_{uvn} z_{uvk}$ in the model.
Given that the number of products per arc is $\mathcal{O}(|V_C|^2)$, the total number of products is $\mathcal{O}(|V_C|^2 |E_0|)$.

\subsection{QUBO model validation}
In order to validate our formulation, the optimal solution of the QUBO model for the 33-node test network was found by classical solvers and this solution was compared with a reference optimal solution found by exhaustive search over the complete spanning trees space of this network.
This search computed the network losses as a direct application of (\ref{eq:optim}) for each spanning tree of the unmodified network model.
Thus, this solution does not depend on any network simplification or any QUBO construct as defined in the previous sections.
The optimal solution found with this method is represented by the set of open links \{(6,7),(8,9),(13,14),(31,32),(24,28)\}.
The remaining network links, and only these ones, belong to the spanning tree corresponding to this solution.
This optimal configuration has a total power loss of 116.379 kW.

As mentioned in Section \ref{sec:power_losses}, the network losses terms in the QUBO model are scaled by a constant factor such that the optimal losses value sits below the constraints penalty threshold of 2.0.
With these losses expressed in kW, this factor was chosen to be 0.01.
Thus, the scaled QUBO value of the optimal solution is 1.16379.

The optimal solution for the QUBO model was found by several MIQP solvers from NEOS Server \cite{neosserver1,neosserver2,neosserver3}: FICO Xpress \cite{ficoxpress} (in 48 seconds), CPLEX \cite{cplex} (in about 5 minutes) and SCIP \cite{SCIP} (in about 23 minutes).
All these solvers returned the expected solution, with the same set of open links and the same QUBO value.

Although this result is not a definitive guarantee of the model correctness, we strongly believe this is the case since it would be extremely unlikely that a mistake on the network losses terms would result in the same computed value for the losses.
Additionally, if the model constraints were insufficient to impose a correct network topology, the optimal solution of the QUBO model would result in a disconnected network in order to unfeasibly decrease the losses value.

\section{Conclusion}\label{section:conclusions}
In this paper, a new QUBO model for the minimum loss reconfiguration problem was proposed.
A comparison with other QUBO formulations of spanning tree reconfiguration problems showed that our formulation is more efficient in terms of the number of variables and interactions.
This result holds not only for the 33-node test network used as an illustrative example, but also for larger networks given the formulation scaling.
This efficiency is given by our general topology constraints -- vertex, edge and cycle constraints -- but it is also a result of the edge lifting transformation, which takes advantage of the linear chains typically found in electrical networks.
Our topology constraints can be used in other QUBO formulations of spanning tree problems in order to obtain a smaller model, particularly for sparse graphs. 

The paper has also shown that the remaining part of our model -- the network losses function -- dominates the QUBO model in terms of the number of variables -- with a quadratic scaling with the network size -- and in terms of the number of interactions -- with a cubic scaling.
Additionally, it was described how edge lifting reduces the number of network losses variables to less than one third.
The number of interactions for the losses function is reduced by the same ratio, given the linear scaling of this number with the number of edges of the reduced graph.

The optimal solution for the QUBO model of the example network was obtained from several solvers.
This solution was found to be the same as the one obtained from exhaustive search with a non-QUBO formulation of the same problem.

With the results obtained for the 33-node network, we expect that our new QUBO formulation will enable the use of quantum annealing or quantum-classical solvers for handling reconfiguration problems on real-world electrical networks with advantage over classical solvers in terms of solution quality and time-to-solution.
This advantage will hopefully become more apparent as quantum annealers and hybrid solvers will continue to improve their performance.

\section*{Acknowledgements}
This work was partially supported by Portuguese national funds through FCT -- Fundação para a Ciência e a Tecnologia with references UIDB/50021/2020 and UIDB/50008/2020.
FFCS thanks the support from Fundação para a Ciência e a Tecnologia and from the European Social Fund through the scholarship SFRH/BD/143402/2019.

\printbibliography

\end{document}